\documentstyle[12pt]{article}
\begin{document}
\begin{center}
{\Large {\bf{TWISTORIAL SUPERPARTICLE}}}

\vspace{0,3cm}
{\Large {\bf{WITH TENSORIAL CENTRAL
CHARGES\footnote{presented at the International
Conference QEDSP 2001, dedicated to the 90th anniversary of
A.I.Akhiezer, October 30 - November 3, 2001}}}}

\vspace{1.5cm}
{\large {\bf S. Fedoruk}}$^\ast$ and {\large {\bf V. G. Zima}}$^{\ast\ast}$

\vspace{0.5cm}
$^\ast$ {\it Ukrainian
Engineering--Pedagogical Academy, \\
61003 Kharkov, 16 Universitetska Str., Ukraine \\
e-mail}: fed@postmaster.co.uk \\

\vspace{0.3cm}
$^{\ast\ast}$ {\it Kharkov National University, \\
61077 Kharkov, 4 Svobody Sq., Ukraine \\
e-mail}: zima@postmaster.co.uk
\end{center}
\vspace{1cm}

\begin{abstract}
A twistorial formulation of the $N=1$ $D=4$ superparticle with
tensorial central charges describing massive and massless cases in
uniform manner is given. The twistors resolve energy-momentum
vector whereas the tensorial central charges are written in term
of spinor Lorentz harmonics. The model makes possible to describe
states preserving all allowed fractions of target-space
supersymmetry. The full analysis of the number of conserved
supersymmetries in models with $N=1$ $D=4$ superalgebra with tensorial
central charges has been carried out.\vspace{1pc}
\end{abstract}
PACS: 11.15.-q, 11.17.+y, 02.40.+m, 11.30.Pb.

\newpage
\section{Introduction}

Some interesting supersymmetric theories admit as scalar central
charges, which are presented in the conventional Poincare
supersymmetry, as well as the nonscallar ones~\cite{GGHT}. In the
supersymmetry algebra tensorial central charges are usually
associated with topological contributions of extended objects. It
is attractive to consider the pure superparticle models having
symmetry of this kind. Such a model was firstly obtained in
massless case with two or three local  $\kappa$--symmetries~\cite{BLS}. We have
constructed the model of massive nonextended superparticle with
central charges~\cite{FZ1} having single  $\kappa$--symmetry, which is equivalent
to conventional spinning (spin $1/2$) particle. In a certain sense
the commuting spinor variables of the model play the role of index
spinor variables~\cite{FZ2}, \cite{FZ3}. Analogous model of massive
superparticle has been formulated in~\cite{DIK}, \cite{BGIK} without Lorentz
invariance. In fact the model~\cite{DIK}, \cite{BGIK} is obtained in particular
case of model~\cite{FZ1} with constant index spinor fixed in non-Lorentz
invariant way.

In recent work~\cite{FZ4} we proposed twistorial
formulation of superparticle with tensorial central charges in
which massive and massless cases are described uniformly. The
model uses both central charge coordinates and auxiliary bosonic
spinor variables simultaneously. In term of the last variables the
energy-momentum vector and the tensorial central charges are
resolved. In the massive case of the proposed model we have
twistorial formulation of massive superparticle with tensorial
central charges preserving $1/4$ or $1/2$ of target-space
supersymmetries. For zero mass this model turns into twistorial
formulation of the massless superparticle with tensorial central
charges~\cite{BLS} in which one or two of target-space supersymmetries
are broken. But the case of massless superparticle with only one
$\kappa$--symmetry is impossible in this formulation.

Our purpose here will
be to give twistorial formulation of $N=1$ $D=4$ superparticle with
tensorial central charges which is able to describe massive and
massless cases in a uniform manner with all allowed possibilities
of the target-space supersymmetry violation. In addition to
twistors we use pair of harmonic spinors by means of which the
tensorial central charges are resolved in Lorentz-covariant way.
But in some cases the choice of corresponding gauges makes
possible to remain only with one type of spinors, for example,
with twistors.

We will in section 2 investigate the $D=4$ $N=1$
superalgebra with tensorial central charges with respect to all
allowed parts of unbroken target-space supersymmetry depending on
the value of the momentum square (particle mass). Section 3
presents the twistor formulation of $D=4$ $N=1$ superparticle with
tensorial central charges. Excepting twistor spinors we use
Lorentz harmonics also. Section 4 describes possible sets of
tensor central charges coefficients in particle action and
corresponding interconnection of harmonic and twistor spinors.
Section 5 contains some comments.

\section{Superalgebra with tensorial central charges}

Generalized central extension of nonextended
$4$--dimensional supersymmetry algebra with Majorana supercharges $Q$ can
be written in the form
$$
\left\{ Q,Q \right\} =2 {\cal Z} \, .
$$
Here ${\cal Z}^T ={\cal Z}$ is the matrix of central charges with a total
of ten real
entries. We have tensorial central charges as coefficients in
decomposition of this matrix
$$
{\cal Z}C=(\gamma^\mu)P_\mu +\frac{i}{2}(\gamma^{\mu\nu})Z_{\mu\nu}
$$
on the basis defined by products of  $\gamma$--matrices. The vector
$P_\mu$ is (in general) a sum of the energy-momentum vector and
a vectorial `string charge'. Namely this vector plays the role of
particle energy-momentum in considering theory. In follows we suppose what
the vector $P_\mu$ satisfies spectral property, i.e. it is positive
time-like or light-like four-vector; $P^2=-m^2$, where $m$ is mass.
Six real charges $Z_{\mu\nu}=-Z_{\nu\mu}$ are related to the
symmetric complex Weyl spin-tensor $Z_{\alpha\beta}=Z_{\beta\alpha}$
in standard way. The spin-tensors $Z_{\alpha\beta}$ and
$\bar Z_{\dot\alpha\dot\beta}=(\overline{Z_{\alpha\beta}})$
represent the self-dual and anti-self-dual parts of the central charge
matrix.

The l.h.s. of the equation for eigenvalues of the matrix ${\cal Z}$,
$$
\Pi(\lambda)\equiv\det({\cal Z}-\lambda)=0
$$
can be written as a polynomial~\cite{GGHT} in $\lambda$,
$$
\Pi(\lambda)=\sum^4_{k=0}\Pi_k \lambda^k \, ,
$$
with
$$
\Pi_0=[(P_0)^2-a]^2 +8bP_0-4c \, ,\quad
\Pi_1=4P_0[(P_0)^2-a]+8b \, ,
$$
$$
\Pi_2=2[3(P_0)^2-a] \, , \quad \Pi_3=4P_0 \, ,\quad \Pi_4=1 \, .
$$
Here
$$
a={\bf E}^2 +{\bf H}^2 +\bf{P}^2 \, ,\quad
b={\bf P}(\bf{E}\times\bf{H}) \, ,\quad
c=|\bf{P}\times\bf{E}|^2+|\bf{P}\times\bf{H}|^2+
|\bf{E}\times\bf{H}|^2
$$
and
$$
E_i=Z^{0i} \, ,\quad H_i=\frac{1}{2}\epsilon_{ijk}Z^{jk}
$$
are electric and magnetic vectors of tensorial central charges.

For the case of massive particle, in the rest frame where $\bf{P}=0$
and $P^0=m$, we can choose the first axis along the vector $\bf{E}$
if it is nonzero. The second axis is embedded in a half-plane defined
by the vector $\bf{H}$ with respect to $\bf{E}$. In the massless
case if $\bf{E}\bf{H}\neq 0$ one can take
$\bf{E}\parallel\bf{H}$ without loss of generality.
If $\bf{E}\bf{H}= 0$ one can choose $\bf{H}=0$ for
$\bf{E}^2-\bf{H}^2>0$  and $\bf{E}=0$ for
$\bf{E}^2-\bf{H}^2<0$.

The massive superparticle with unique preserved SUSY
($\Pi_0 =0$, $\Pi_1\neq 0$) is obtained only if
$$
(m^2 -\bf{E}^2-\bf{H}^2)^2=
4|\bf{E}\times\bf{H}|^2\neq 0 \, .
$$
Up to rotations the variety of such configurations is
characterised by positive modulus of non-collinear vectors $\bf{E}$
and $\bf{H}$ and the angle $\vartheta$ between them.
At the boundary of this region of parameters when
$m^2 =\bf{E}^2+\bf{H}^2\neq 0$ with collinear $\bf{E}$ and
$\textbf{H}$ we have two preserved SUSYs ($\Pi_0 =\Pi_1=0$, $\Pi_2\neq 0$).
Conditions for preserving more than two SUSYs ($\Pi_0 =\Pi_1=\Pi_2=0$) in
massive case are contradictory.

In massless case there are two
types of configurations preserving unique SUSY. One of them takes
place for the collinear ${\bf E}$ and ${\bf H}$ with non-collinear
to them ${\bf P}$ if ${\bf E}^2 +{\bf H}^2 =4{\bf H}^2 \sin^2\varphi\neq0$
where $\varphi$ is the angle between ${\bf P}$ and ${\bf E}$. Another
one consists in mutual orthogonality of three vectors  ${\bf E}$, ${\bf H}$
and ${\bf P}$ forming right triple, ${\bf P}{\bf E}{\bf H}>0$, and
equality of two modulus, $|{\bf E}|=|{\bf H}|\neq P^0$. At the boundary
of the first of these regions we have the odd sector of conventional
massless superparticle without central charges (${\bf E}={\bf H}=0$)
preserving two SUSYs. We have two preserved SUSYs also if three mutually
orthogonal vectors ${\bf E}$, ${\bf H}$ and ${\bf P}$ form left triple
${\bf P}{\bf E}{\bf H}<0$ and two of them have equal lengths
$$
|{\bf E}|=|{\bf H}|\neq P^0 \, .
$$
At the boundary of this region if mutually orthogonal vectors form
left triple and all of them have equal modulus
$$
|{\bf E}|=|{\bf H}|= P^0 \, ,
$$
then three SUSYs are preserved. All the SUSYs cannot be preserved
because of $\Pi_3=4P_0\neq 0$.

In the Table {\bf 1} all possible preservations of fractions
of target-space supersymmetry for massless ($m=0$) and massive ($m\neq 0$)
$D=4$ $N=1$ superparticle with tensorial central charges are given.
Except the case of massless superparticle with $1/4$ unbroken SUSY,
in Table 1 it is mention the full set of necessary conditions.
For massless superparticle ($m=0$) with $1/4$ unbroken SUSY these conditions are
complicated. We give required conditions at orthogonal vectors ${\bf E}$,
${\bf H}$ to ${\bf P}$,
$$
{\bf P}{\bf E}={\bf P}{\bf H}=0 \, .
$$
Here $\varepsilon=+1$ at ${\bf P}{\bf E}{\bf H}>0$ or
$\varepsilon=-1$ at ${\bf P}{\bf E}{\bf H}<0$.

\vspace{1cm}
\begin{center}
\begin{tabular}{|c|c|c|c|}
\hline
   & ${\bf 1/4}$ & ${\bf 1/2}$ & ${\bf 3/4}$ \\
   \hline
   & & \multicolumn{2}{|c|}{\,\,}\\
& ${\bf E}\times{\bf H}=0$, ${\bf E}^2+{\bf H}^2=4P_0^2$ &
\multicolumn{2}{|c|}{${\bf E}{\bf H}=0$,}\\
& (${\bf P}{\bf E}={\bf P}{\bf H}=0$)&
\multicolumn{2}{|c|}{${\bf E}^2-{\bf H}^2=0$,}\\
$m=0$ & or & \multicolumn{2}{|c|}{${\bf P}{\bf E}={\bf P}{\bf H}=0$,}\\
& ${\bf E}{\bf H}=0$, $|{\bf E}|-\varepsilon |{\bf H}|=\pm P^0$ &
\multicolumn{2}{|c|}{\,\,}\\
& (${\bf P}{\bf E}={\bf P}{\bf H}=0$)
& ${\bf E}^2+{\bf H}^2\neq 2P_0^2$ & ${\bf E}^2+{\bf H}^2= 2P_0^2$\\
& & & \\\hline
& & & \\
$m\neq 0$ & $(m^2 -{\bf E}^2-{\bf H}^2)^2=$
 & ${\bf E}\times{\bf H}=0$,
& no \\
& $=4|{\bf E}\times{\bf H}|^2\neq 0$& ${\bf E}^2+{\bf H}^2=m^2$ & \\
& & & \\\hline
\end{tabular}
\end{center}
\vspace{0.5cm}

{\bf Table 1.} The conditions on `electric' and `magnetic' vectors
of tensorial central charges and energy-momentum vector for states of
$N=1$ $D=4$ massless ($m=0$) and massive ($m\neq 0$) superparticles
preserving fractions $1/4$, $1/2$ and $3/4$ of target-space supersymmetry

\vspace{0.5cm}

In the following we present the twistorial
formulation of the superparticle with tensorial central charges in
which all allowed cases of supersymmetry violation can be realized
both for massive superparticle and massless one.

\section{Lagrangian of twistorial superparticle with tensorial central
charges}

A trajectory of superparticle is parameterized by the
usual superspace coordinates $x^\mu$, $\theta^\alpha$,
$\bar\theta^{\dot\alpha}$ and tensorial central charge
coordinates  $y^{\alpha\beta}=y^{\beta\alpha}$,
$\bar y^{\dot\alpha\dot\beta}=(\overline{y^{\alpha\beta}})$. For
description of superparticle energy-momentum
we use the pair of bosonic spinors (bitwistor) $v_\alpha{}^a$,
$\bar v_{\dot\alpha a}=(\overline{v_\alpha{}^a})$, $a=1,2$. The tensor
central charges are written in terms of even spinor variables
$u_\alpha{}^a$, $\bar u_{\dot\alpha a}=(\overline{u_\alpha{}^a})$,
$a=1,2$. We use $D=4$ Weyl spinor and  $\sigma$--matrices conventions
of~\cite{WB}
where the metric tensor $\eta_{\mu\nu}$ has mostly plus and
$\sigma_{\alpha\dot\beta}^{(\mu}\tilde\sigma^{\nu)\dot\beta\gamma}=
-\delta_\alpha^\gamma$. The indices $a$, $b$, $c$ …  which are
carried by the spinors $v_\alpha{}^a$,
$\bar v_{\dot\alpha a}$ as well as $u_\alpha{}^a$,
$\bar u_{\dot\alpha a}$, are
raised and lowered as ${\bf SU(2)}$ ones.

For description of the superparticle, both massless and massive,
with tensorial central charges we take the action $S=\int d\tau L$
with Lagrangian~\cite{FZ4} in twistor-like form
\begin{equation}
L=P_\mu\Pi^\mu_\tau +Z_{\alpha\beta}\Pi^{\alpha\beta}_\tau +
\bar Z_{\dot\alpha\dot\beta}\bar\Pi^{\dot\alpha\dot\beta}_\tau -
\lambda_v h_v - \bar\lambda_v \bar h_v -
\lambda_u h_u - \bar\lambda_u \bar h_u \, .
\end{equation}
Here the one-forms
$$
\Pi^{\mu}\equiv d\tau\Pi^{\nu}_{\tau}= d x^\mu -
id\theta\sigma^\mu\bar\theta +i\theta\sigma^\mu d\bar\theta \, ,
$$
$$
\Pi^{\alpha\beta}\equiv d\tau\Pi^{\alpha\beta}_{\tau}=
d y^{\alpha\beta} +i\theta^{(\alpha}d\theta^{\beta)} \, ,
$$
$$
\bar\Pi^{\dot\alpha\dot\beta}\equiv
d\tau\bar\Pi^{\dot\alpha\dot\beta}_{\tau}=
d \bar y^{\dot\alpha\dot\beta} +
i\bar\theta^{(\dot\alpha}d\bar\theta^{\dot\beta)}
$$
are invariant under global supersymmetry transformations
$$
\delta\theta^\alpha=\epsilon^\alpha \, ,\quad
\delta\bar\theta^{\dot\alpha}=\bar\epsilon^{\dot\alpha} \, ;
$$
$$
\delta x^\mu =i\theta\sigma^\mu\delta\bar\theta -
i\delta\theta\sigma^\mu\bar\theta \, ;
$$
$$
\delta y^{\alpha\beta}=i\theta^{(\alpha}\delta\theta^{\beta)} \, ,\quad
\delta \bar y^{\dot\alpha\dot\beta}=
i\bar\theta^{(\dot\alpha}\delta\bar\theta^{\dot\beta)}
$$
acting in the extended superspace parameterised by the usual
superspace coordinates $x^\mu$, $\theta^\alpha$,
$\bar\theta^{\dot\alpha}$ and by the tensorial central charge
coordinates  $y^{\alpha\beta}=y^{\beta\alpha}$,
$\bar y^{\dot\alpha\dot\beta}$.

In the action the quantities $P_\mu$, $Z_{\alpha\beta}$,
$\bar Z_{\dot\alpha\dot\beta}=(\overline{Z_{\alpha\beta}})$ which play
the role of the momenta for $x^\mu$, $y^{\alpha\beta}$,
$\bar y^{\dot\alpha\dot\beta}$ are taken in the form
\begin{equation}
P_{\alpha\dot\beta}=P_\mu \sigma^\mu_{\alpha\dot\beta}=
v_\alpha{}^a \bar v_{\dot\alpha a} \, ,
\end{equation}
\begin{equation}
Z_{\alpha\beta} =u_\alpha{}^a u_\beta{}^b C_{ab} \, ,\quad
\bar Z_{\dot\alpha\dot\beta}=
\bar u_{\dot\alpha a} \bar u_{\dot\beta b}\bar C^{ab}
\end{equation}
where $C_{ab}$, $\bar C^{ab}=(\overline{C_{ab}})$ are symmetric
constants. Thus they are resolved in
terms of bosonic Weyl spinors $v_\alpha{}^a$,
$\bar v_{\dot\alpha a}$ and $u_\alpha{}^a$,
$\bar u_{\dot\alpha a}$.

The last terms in Lagrangian (1) are the sum of the
kinematic constraints for the even spinors
\begin{equation}
h_v \equiv v^{\alpha a} v_{\alpha a} -2m\approx 0 \, ,\quad
\bar h_v \equiv \bar v_{\dot \alpha a} \bar v^{\dot\alpha a} -
2m \approx 0 \, ;
\end{equation}
\begin{equation}
h_u \equiv u^{\alpha a} u_{\alpha a} -2\approx 0 \, ,\quad
\bar h_u \equiv \bar u_{\dot \alpha a} \bar u^{\dot\alpha a} -
2 \approx 0
\end{equation}
with Lagrange multipliers.

Due to the constraints (4) which are

equivalent to
$$
v^{\alpha a} v_{\alpha b} =m \delta^a_b \, ,\quad
\bar v_{\dot \alpha a} \bar v^{\dot\alpha b} =m \delta_a^b
$$
we have $\det (v_{\alpha}{}^a)=m$ and
$$
P^2 \equiv P^\mu P_\mu =-m^2 \, .
$$
Thus the constant $|m|$ plays the role of the mass.

In both cases, massless and massive, we have twistorial
resolution~\cite{P} of energy-momentum vector in term of bosonic spinors.

In the massless case ($m=0$) the spinors $v_{\alpha}{}^1$ and
$v_{\alpha}{}^2$ are proportional to each other
$v_{\alpha}{}^1 \propto v_{\alpha}{}^2$ as
the consequence of the constraint (4) $v^{\alpha 1} v_{\alpha}{}^2 =0$.

The constraints (5) are equivalent to
$$
u^{\alpha a} u_{\alpha b} = \delta^a_b \, ,\quad
\bar u_{\dot \alpha a} \bar u^{\dot\alpha b} = \delta_a^b
$$
and they give $\det (u_{\alpha}{}^a)=1$. Thus the spinors $u_\alpha{}^a$,
$\bar u_{\dot\alpha a}$ play the role of harmonic
variables~\cite{GIKOS}-\cite{FZ5} parametrizing an appropriate homogeneous
subspace of the Lorentz group.

In addition to kinematical constraints (4), (5) the bosonic
spinor variables of the model are satisfied to constraints
on spinor momenta $p_v\approx 0$, $p_u \approx 0$ and c.c. A part of
these constraints conjugate to constraints (2-5) and are second
class constraints. The rest of the constraints on spinor momenta,
which conserve the constraints (2), (3), are first class
constraints and correspond to stability subgroup of Lorentz group
acting on indices $a$, $b$, ... The choices of gauge allow to connect
harmonic spinors and twistor  ones for certain cases of the
central charges and particle mass.

\section{Superparticle states preserving arbitrary fractions
of target-space supersymmetry}

The expressions (3) of central charges containing three complex (six
real) constants are collected in $C_{ab}$, $\bar C^{ab}$.
The symmetric matrices are expanded in symmetric Pauli matrices
$$
(\sigma_i)_{ab} =\epsilon_{ac}(\sigma_i)_a{}^c \, ,\quad
(\sigma_i)^{ab} =\epsilon^{ac}(\sigma_i)_c{}^b \, ,
$$
where $(\sigma_i)_a{}^b$, $i=1,2,3$ are usual Pauli matrices. Thus
$$
C_{ab}=C_{i}(\sigma_i)_{ab} \, ,\quad
\bar C^{ab}=-\bar C_i(\sigma_i)^{ab}
$$
where $\bar C_i=(\overline{C_{i}})$. We obtain directly
$$
C_{ab}C^{bc}=-C_{i}C_{i}\delta_a^c =
({\bf E}^2 -{\bf H}^2+2i{\bf E}{\bf H})\delta_a^c \, ,
$$
$$
C_{ab}\bar C^{ab}=2C_{i}\bar C_i =2({\bf E}^2+{\bf H}^2) \, ,
$$
where real $3$--vectors ${\bf E}$ and ${\bf H}$ are defined by
the equality
$$
{\bf C}=i({\bf E}+i{\bf H}) \, .
$$

The vectors ${\bf E}$ and ${\bf H}$ are `electric' and `magnetic'
vectors of tensorial central charges, which was used in Section 2.
But now they are determined with respect to the basis of Lorentz
harmonics $u_\alpha{}^a$, $\bar u_{\dot\alpha a}$.
In this basis the components of energy-momentum are
$$
P^{(0)}=\frac{1}{2}P_a{}^a \, ,\quad
P^{(i)}=\frac{1}{2}P_a{}^b (\sigma_i)_b{}^a
$$
where matrices $P_a{}^b$ and $P_{\alpha\dot\beta}$ are connected
by the Lorentz transformation generated by harmonic matrix
$u_\alpha{}^a$, i.e.
$$
P_{\alpha\dot\beta}=u_\alpha{}^a \bar u_{\dot\beta b}P_a{}^b \, .
$$
Thus the condition
\begin{equation}
u_\alpha{}^a \bar u_{\dot\beta b}P_a{}^b =
v_\alpha{}^a \bar v_{\dot\beta a}
\end{equation}
connects bitwistor representation of the energy-momentum and its
representation in harmonic basis.

For given energy-momentum vector ${\bf P}$
the fraction of preserving supersymmetry is determined by the
choice of `electric' ${\bf E}$ and `magnetic' ${\bf H}$ vectors.
Since central charges are written in terms of harmonics,
any choice of ${\bf E}$ and ${\bf H}$
does not lead to violation of Lorentz invariance. In the following
it is convenient to make the analysis in the standard momentum
frame.

\subsection{Massive ($m\neq 0$) case}

In the frame of standard momentum $P^{(0)}=m$ , $P^{(i)}=0$ the
expression (6) gives
$$
m u_\alpha{}^a \bar u_{\dot\beta a} =
v_\alpha{}^a \bar v_{\dot\beta a}
$$
Hence the harmonic spinors $u_\alpha{}^a$ and twistor ones
$v_\alpha{}^a$ are identical up to unitary transformations acting
on index $a$. Without loss of generality we can take
$$
m^{1/2}u_\alpha{}^a =v_\alpha{}^a \, .
$$
This identification can be obtained by gauge fixing of the
harmonic degrees of freedom, which are pure gauge ones in the
initial action (1). As result, we derive the model of
superparticle with only twistor spinors (or only with Lorentz
harmonics). Such model was considered in~\cite{FZ4} and gives all
possible cases of target-space supersymmetry preserving for
massive superparticle.

The case with $1/2$ unbroken SUSY is reached if
central charge coefficients satisfy the conditions
\begin{equation}
C^{ab}\bar C_{ab}=2m^2 \, ,\quad
C^{ab} C_{ab} \bar C^{cd}\bar C_{cd}=4m^4
\end{equation}
which are equivalent to
$$
{\bf E}^2+{\bf H}^2 =m^2 \, ,\quad
{\bf E}\times {\bf H}=0 \, .
$$
As it been obtained in~\cite{FZ4}, the conditions (7) give the unitary
condition on the coefficient matrix of central charges
$$
C^{ab}\bar C_{bc}=m^2\delta^a_c
$$
Without loss of generality we can take diagonal matrix $C^{ab}$ with
$m$ multiplied on phase multipliers on diagonal of them. This
corresponds to vectors ${\bf E}$ and ${\bf H}$ which both set
in plane $XY$.

The case with $1/4$ unbroken SUSY is obtained for
${\bf E}\times {\bf H}\neq0$ when
$$
|{\bf E}^2+{\bf H}^2 -m^2|=2|{\bf E}\times {\bf H}|
$$
is satisfied. Here only one from two elements of diagonal matrix $C_{ab}$
has the modulus equal to $m$.

\subsection{Massless ($m=0$) case}

In the standard energy-momentum frame for massless particle
$$
P^\mu =(P^{(0)},0,0,P^{(0)})
$$
and
$$
P_a{}^b =2P^{(0)}(\sigma_+)_a{}^b
$$
where $\sigma_+=(1_2 +\sigma_3)/2$. The twistor spinors,
which are proportional each other in massless case can be taken equal
$$
v_\alpha\equiv v_\alpha{}^1 =v_\alpha{}^2 \, .
$$
Then the expression (6) give
$$
2P^{(0)}u_\alpha{}^1 \bar u_{\dot\beta 1} =
2v_\alpha \bar v_{\dot\beta} \, .
$$
Therefore one harmonic spinor $u_\alpha{}^1$ is expressed in form
twistor one $v_\alpha$
up to phase transformation. Thus we can remain with twistor spinor
$v_\alpha$ and one harmonic spinor $u_\alpha{}^2$.
The second harmonic spinor is obtained
directly from twistor one (for example,
$u_\alpha{}^1=(P^{(0)})^{-1/2}v_\alpha$) if it is necessary.

In terms of vectors ${\bf E}$ and ${\bf H}$ the matrix   has the following expression
$$
C_{ab}=\left(
\begin{array}{cc}
H_1 -E_2 -i(H_2 +E_1) & -H_3 +iE_3 \\
-H_3 +iE_3 & -H_1 -E_2 -i(H_2 -E_1)
\end{array}
\right)
$$

In case of $1/2$ and $3/4$ unbroken SUSY vectors ${\bf E}$ and ${\bf H}$
are orthogonal to each other and to vector ${\bf P}$  (see Table 1)
which along third axis.
Therefore $E_3=H_3=0$ and matrix $C_{ab}$ is diagonal. In these cases
($1/2$ and $3/4$ unbroken SUSY) vectors ${\bf E}$ and ${\bf H}$ have
equal lengths $|{\bf E}|=|{\bf H}|\equiv V$ and the
nonzero elements of the matrix $C_{ab}$ are  $C_{11}=2Ve^{-i\phi}$
in case ${\bf P}{\bf E}{\bf H}>0$ or $C_{22}=-2Ve^{i\phi}$ in case
${\bf P}{\bf E}{\bf H}<0$ where $\phi$ is angle between first axis
and vector ${\bf H}$. In case of $1/2$
unbroken SUSY it is just $V\neq P^{(0)}$ whereas in case of $3/4$
unbroken SUSY it
is fulfilled $V= P^{(0)}$. The case with ${\bf P}{\bf E}{\bf H}>0$
and nonzero only $C_{11}\neq 0$ correspond to
twistor model of the superparticle with tensorial central charges
considered in~\cite{BLS}.

For massless particle the states which
preserving $1/4$ SUSY are realised at $|{\bf E}|\neq |{\bf H}|$ when
${\bf E}{\bf H}=0$. At these conditions
the matrix $C_{ab}$ has two nonzero diagonal elements. Namely
$$
C_{11}=(|{\bf E}|+|{\bf H}|)e^{-i\phi} \, ,\quad
C_{22}=(|{\bf E}|-|{\bf H}|)e^{i\phi}
$$
in case ${\bf P}{\bf E}{\bf H}>0$ or
$$
C_{11}=-(|{\bf E}|-|{\bf H}|)e^{-i\phi} \, ,\quad
C_{22}=-(|{\bf E}|+|{\bf H}|)e^{i\phi}
$$
in case  ${\bf P}{\bf E}{\bf H}<0$. Thus the case with only one
$\kappa$--symmetry is realised for
massless superparticle if two spinors are presented in model. The
using of one twistor spinor as in~\cite{BLS} is insufficient for
realization of $1/4$ unbroken SUSY. The more strong violation of
target-space supersymmetry requires using more numbers of the
spinors represented central charges of supersymmetry algebra~\cite{BAIL}.

\section{Conclusion}

In present paper we construct the model of $N=1$ $D=4$
superparticle with tensorial central charges. By corresponding
choice of tensorial central charges we obtain all possibilities of
violated target-space supersymmetry. It should be noted that these
choices are fulfilled in Lorentz-covariant way due to using
Lorentz harmonics in model. The twistor formulation of the massive
superparticle is required to use of pair of twistors. For massless
superparticle the cases with $3/4$ and $1/2$ unbroken SUSY are realized by
using only one twistor spinor, the energy-momentum and the
tensorial central charges are resolved in form of which. But for
realization of only $1/4$ unbroken SUSY it is necessary to use two
spinors in model, one from which is twistor whereas second used
spinor can be harmonic one.

This work was supported in part by
INTAS Grant INTAS-2000-254 and by the Ukrainian National Found of
Fundamental Researches under the Project N 02.07/383. We would
like to thank I.A.~Bandos, A.~Frydryszak, E.A.~Ivanov, S.O.~Krivonos,
J.~Lukierski, A.J.~Nurmagambetov and D.P.~Sorokin for
interest to the work and for many useful discussions.

\end{document}